 \documentclass[final,1p,times]{elsarticle}
 \usepackage{mathrsfs}
\usepackage{amsmath}
\usepackage{graphics}
\usepackage{txfonts}
\usepackage{setspace}
\usepackage{amsfonts,amssymb,graphicx,mathrsfs,epsfig,subfigure}
\usepackage[dvipdfm,colorlinks,linkcolor=blue,citecolor=blue]{hyperref}
\usepackage{multirow}
\usepackage{booktabs}
\usepackage{fancyhdr,times}

\usepackage{lineno}

\biboptions{sort&compress}

\newproof{proof}{Proof}
\newdefinition{remark}{Remark}
\begin{document}
%  \linenumbers
\begin{frontmatter}

\title{\Large \bf Bistability in a SIRS model with general nonmonotone and saturated incidence rate  \tnoteref{t1}}

 \tnotetext[t1]{This work is supported by NSFC (No.U1604180), Key Scientific and Technological
Research Projects in Henan Province (No.192102310089), Foundation of
Henan Educational Committee (No.19A110009) and Grant of
Bioinformatics Center of Henan University (No.2019YLXKJC02).}
\author[els]{Shaoli Wang \corref{cor1}}
\ead{wslheda@163.com }
\author[els]{Xiyan Bai  }
\author[wlu]{Fei Xu} \ead{fxu.feixu@gmail.com}

 \cortext[cor1]{Corresponding author.}

\address[els]{School of Mathematics and Statistics, Henan University, Kaifeng 475001, Henan, PR China }
\address[wlu]{Department of Mathematics, Wilfrid Laurier University, Waterloo, Ontario, Canada \ N2L 3C5}

\begin{abstract}
% Text of abstract

\begin{spacing}{1.0}
 %%行间距变为single-space

Recently, Lu et al. [J. Differential Equations, 267 (2019)
1859--1898] studied a susceptible-infectious-recovered (SIRS)
epidemic model with generalized nonmonotone and saturated incidence
rate $\frac{kSI^{2}}{1+\beta I+\alpha I^{2}}$. With the emergence of
a new infectious disease,  the infection function  increases to a
maximum, which is followed by a decrease caused by  psychological
effect. The infection function   eventually approaches  a saturation
level due to crowding effect. Then they analyzed the dynamic
behaviors of the reduced system. In this paper, we show that the
unreduced system has saddle-node bifurcation and displays bistable
behavior, which is a new phenomenon in epidemic dynamics and
different from the backward bifurcation behavior. We obtain the
critical thresholds that characterize the dynamical behaviors of the
model. We find with surprise that the system always admits a disease
free equilibrium which
 is always asymptotically stable.

 \end{spacing}

\end{abstract}


\end{frontmatter}

\section{Introduction}

The spread of infectious diseases causes a crisis in public health
and threats the survival of human population. Controlling the
transmission of disease has been a major concern of the society. In
the literature, researchers proposed mathematical models to
characterize the spread and control of epidemics. As early as 1927,
 Kermazk--McKendrick model was proposed to study the infectious diseases using dynamics method. In the past 30 years, the dynamics of infectious diseases have been extensively studied.
 A large number of mathematical models have been proposed to model a variety of infectious diseases including SIR, SIS SIRS, SEIR, SEIS and SEIRS models \cite{Hethcote,Marcos,Hethcote1}.
  As a classic epidemic model, SIRS was investigated by Kermazk and McKendrick \cite{McKendrick}. The model assumes that a recovered individual will obtain temporary immunity against the disease.
  Since the immunity is not permanent, the recovered individual will lose the immunity and become susceptible after a period of time. In the SIRS model, $S$ represents number of susceptible individuals,
   $I$ represents number of infectious individuals, and $R$ represents the number of recovered individuals.

In order to incorporate the effect of behavioral changes, nonlinear
incidence function of the form   $H_{1}(S,I)=\lambda S^{p}I^{q}$ and
a more general form
  $H_{2}(S,I)=\frac{\lambda S^{p}I^{q}}{1+vI^{p-1}}$ were developed and studied by Liu, Levin, and Iwasa in \cite{Liu,Liu2}.
  The authors investigated the behaviors of the epidemic model with the nonlinear incidence rate $H_{1} (S,I)$
   and found that the dynamical behaviours of the model are determined mainly by $p$ and $\lambda$, and secondarily by $q$. Their investigation explained how such a nonlinearity arise.
Based on the work of Liu \cite{Liu}, Hethcote and van den Driessche
\cite{Hethcote2} used a nonlinear incidence rate of the form
$H(S,I)=\frac{kSI^{p}}{1+\alpha I^{q}}$, where $kI^{p}$ represents
the infection force of the disease, $1+\alpha I^{q}$ is a
description of the suppression effect from the behavioral changes of
susceptible individuals when the infective population increases,
$p>0,\,p>0$ and $α≥0$. The authors studied the number of
disease-free and endemic equilibria of an SEIRS epidemic model for
$p=q$ and $p>q$, and considered their stability.  Ruan and Wang
\cite{Ruan1} studied the global dynamics of an epidemic model with
vital dynamics and nonlinear incidence rate of saturated mass action
$\frac{kSI^{2}}{1+\alpha I^{2}}$ and Tang et al. \cite{Ruan2}
investigated  the coexistence of a limit cycle and a homoclinic loop
in this model. Li and Teng \cite{Li} considered an SIRS epidemic
model with a more generalized non-monotone incidence:
$\frac{kI^{p}}{1+I^{q}}$ with $0< p < q$. The authors found that for
different values of $p$, the model displays different dynamic
behaviors.

Lu et al. \cite{Hung} provided a more reasonable incidence function
$\frac{kSI^{2}}{1+\beta I+\alpha I^{2}}$, which first increases to a
maximum when a new infectious disease emerges or an old infectious
disease reemerges, then decreases due to psychological effect, and
eventually tends to a saturation level due to crowding effect. On
the other hand, in some specific infectious diseases, the incidence
rate may not be monotonic or non-monotonic alone, a more general
incidence rate may have a combination of monotonicity,
nonmonotonicity and saturation properties. Then they carried out
dynamical analysis of a SIRS epidemic model with the generalized
nonmonotone and saturated incidence rate.

In this paper, we will investigate the SIRS model
$$\begin
{array}{l l} \frac{dS}{dt}=b-dS-\frac{kSI^{2}}{1+\beta I+\alpha I^{2}}+\delta R, \\
\frac{dI}{dt}=\frac{kSI^{2}}{1+\beta I+\alpha I^{2}}-(\mu+d)I, \\
\frac{dR}{dt}=\mu I-(\delta+d)R,
  \end {array}
   \eqno(1.1) $$
where $S,I,R$ are the numbers of susceptible, infective, and
recovered individuals at time $t$. In the model, all parameters take
positive values.  Here, $b$ is the natural birth rate. The natural
decay rate is $d$, the disease related death rate is $\mu$,
 the infection rate is $k$, and $\alpha$ and $\beta$ are parameters for inhibitory effect. Recovered individuals lose immunity and move into susceptible compartment $S$ at rate $\delta$.

\section{Equilibria and thresholds}

Denote $$ R_{0}=\frac{bk}{\beta d(\mu+d)},$$
$$R_{c}=R_{0}-\frac{2}{\beta}\sqrt{\alpha+\frac{k}{d}(1-\frac{
\mu\delta}{(\mu+d)(\delta+d)})},$$ and $$
R_{cc}=R_{0}+\frac{2}{\beta}\sqrt{\alpha+\frac{k}{d}(1-\frac{
\mu\delta}{(\mu+d)(\delta+d)})}.$$ It is easy to see that
$R_{c}<R_{0}<R_{cc}$.

(i) System (1.1) always has a disease-free equilibrium
$E_{0}=(\frac{b}{d}, 0,0)$.

(ii)  To obtain the positive equilibria of system (1.1), we solve
the following equations:

$$\begin
{array}{l l} b-dS-\frac{kSI^{2}}{1+\beta I+\alpha I^{2}}+\delta R=0, \\
\frac{kSI}{1+\beta I+\alpha I^{2}}-(\mu+d)=0, \\
\mu I-(\delta+d)R=0.\
  \end {array}
   \eqno(2.1) $$
From the third equation of system (2.1) we have $R=\frac{\mu
I}{d+\delta}.$ Solving equation
$$b-dS-\frac{kSI^{2}}{1+\beta I+\alpha I^{2}}+\delta R=0,$$
we have
$$S=\frac{b-(\mu+d-\frac{\delta\mu}{\delta+d})I}{d}.$$
Substituting $S$ and $R$ into above equation yields s

$$AI^{2}+BI+1=0, \eqno(2.2) $$
where $$A=\alpha+\frac{k}{d}(1-\frac{\mu\delta}{
(\mu+d)(\delta+d)}),$$ $$B=\beta(1-R_{0}).$$ We have
$$\Delta=B^{2}-4A. $$
If $\Delta>0$, then $R_{cc}<1$ or  $R_{c}>1.$ If $B<0$, then
$R_{0}>1$. Thus, we should just consider the case $R_{c}>1$. In this
case, equation (2.2) has two positive roots:
$$ I_{\pm}^{*}=\frac{-B\pm\sqrt{\Delta}}{2A}.$$

\noindent{\bf Theorem 2.1}\hspace{0.1cm} (i) System (1.1) always has
a disease-free equilibrium $E_{0};$

(ii) If $R_{c}>1$, system (1.1) also has two positive equilibria
$E_{+}^{*} (S_{+}^{*}, I_{+}^{*}, R_{+}^{*}), E_{-}^{*} (S_{-}^{*},
I_{-}^{*}, R_{-}^{*}),$ where

$$S_{+}^{*}=\frac{b-(\mu+d-\frac{\delta\mu}{\delta+d})I_{+}^{*}}{d}, I_{+}^{*}=\frac{-B+\sqrt{\Delta}}{2A}, R_{+}^{*}=\frac{\mu
I_{+}^{*}}{d+\delta}, $$
$$S_{-}^{*}=\frac{b-(\mu+d-\frac{\delta\mu}{\delta+d})I_{-}^{*}}{d}, I_{-}^{*}=\frac{-B-\sqrt{\Delta}}{2A}, R_{-}^{*}=\frac{\mu
I_{-}^{*}}{d+\delta}. $$

%%%20191122
The existence of positive equilibria are summarized in Table $1$.

\begin{table*}[ht]
\caption{The existence of the positive equilibria  of system (1.1)}
\begin{center}
\begin{tabular}{|l|ll|l|}
\hline   & $R_{c}<1$  &$R_{c}>1$
\\\hline
$E_{0}$                   &exist     &exist     \\
 $E_{+}^{*}$      &  --- & exist     \\
 $E_{-}^{*}$        &  --- &  exist    \\
\hline
\end{tabular}
\end{center}
\end{table*}

\section{Stability analysis}

Let $\tilde{E}$ be any arbitrary equilibrium of system (1.1). The
Jacobian matrix associated with  system (1.1) is

$$\mathscr{J}_{\tilde{E}}=\left[
\begin{array}{cccc}
 -d-\frac{k\widetilde{I}^{2}}{1+\beta \widetilde{I}+\alpha \widetilde{I}^{2}}     &-\frac{k\widetilde{S}\widetilde{I}(2+\beta \widetilde{I})}{(1+\beta \widetilde{I}+\alpha \widetilde{I}^{2})^{2}}       &\delta \\
\frac{k\widetilde{I}^{2}}{1+\beta \widetilde{I}+\alpha \widetilde{I}^{2}}        &\frac{k\widetilde{S}\widetilde{I}(2+\beta \widetilde{I})}{(1+\beta \widetilde{I}+\alpha \widetilde{I}^{2})^{2}}-(\mu+d)   & 0  \\
0    &\mu    &-(\delta+d)  \\
\end{array}
\right].$$ The characteristic equation of  system (1.1) at
$\tilde{E}$ is  $\left|\lambda I-\mathscr{J}_{\tilde{E}} \right|=0.$

\textbf{3.1. Stability analysis of  the disease-free equilibrium }

\noindent{\bf Theorem 3.1 }\hspace{0.1cm} The disease-free
equilibrium $E_{1}$ of system (1.1) is always locally asymptotically
stable.

\noindent{\bf Proof.}

The characteristic equation of the linearized system of (1.1) at the
disease-free equilibrium $E_{0}$ is obtained as

$$(\lambda +d)(\lambda +d+\mu)(\lambda+d+\delta)=0.$$

The characteristic polynomial  has three roots $-d$, $-(\mu+d)$  and
$-(\delta+d)$. Since the three roots are all negative,  the
disease-free equilibrium $E_{0}$ of system (1.1) is locally
asymptotically stable. This completes the proof of Theorem 3.1\qed

\textbf{3.2. Stability analysis of positive equilibria}

 \noindent{\bf Theorem 3.2}\hspace{0.1cm} If $R_{c}>1$,
$a_{1}a_{2}-a_{3}>0,$ and $a_{2}>0$, system (1.1) has two positive
equilibria $E_{+}^{*}$ and $E_{-}^{*}$, where $E_{+}^{*}$ is a
locally asymptotically stable node and $E_{-}^{*}$ is an unstable
saddle.

\noindent{\bf Proof.}

Denote  an arbitrary positive equilibrium of system (1.1) as
$E^{*}$. The characteristic equation of the system (1.1) at the
arbitrary positive equilibrium $E^{*}$ is obtained as

$$\lambda^{3}+a_{1}\lambda^{2}+a_{2}\lambda+a_{3}=0,$$
where
$$\begin {array}{lll}
a_{1}=\mu+\delta+3d+\frac{k(I^{*})^{2}}{1+\beta I^{*}+\alpha (I^{*})^{2}}-\frac{(\mu+d)(2+\beta I^{*})}{1+\beta I+\alpha (I^{*})^{2}},\\
a_{2}=(\delta+2d)(\mu+d)+d(\delta+d)+(\mu+\delta+2d)\frac{k(I^{*})^{2}}{1+\beta I^{*}+\alpha (I^{*})^{2}}-(\delta+2d)\frac{(\mu+d)(2+\beta I^{*})}{1+\beta I^{*}+\alpha (I^{*})^{2}},\\
a_{3}=d(\delta+d)(\mu+d)+(\delta+d)(\mu+d)\frac{k(I^{*})^{2}}{1+\beta I^{*}+\alpha (I^{*})^{2}}-d(\delta+d)\frac{(\mu+d)(2+\beta I^{*})}{1+\beta I^{*}+\alpha (I^{*})^{2}}-\mu \delta \frac{k(I^{*})^{2}}{1+\beta I^{*}+\alpha (I^{*})^{2}}.\\
\end {array}$$

(i) For equilibrium $E_{+}^{*},$ we have
$$\begin {array}{lll}
d\alpha (I_{+}^{*})^{2}+k(I_{+}^{*})^{2}-d-\mu \delta \frac{k(I_{+}^{*})^{2}}{(d+\delta)(d+\mu)},\\
=d(\alpha+\frac{k}{d}(1-\frac{\mu\delta}{(d+\delta)(d+\mu)}))(I_{+}^{*})^{2}-d,\\
=d(\alpha+\frac{k}{d}(1-\frac{\mu\delta}{(d+\delta)(d+\mu)}))\frac{(\beta(R_{0}-1)+\sqrt{\Delta})^{2}}{4A^{2}}-d,\\
=\frac{d(\beta(R_{0}-1)+\sqrt{\Delta})^{2}}{4A}-d.\\
\end {array}$$
It follows from
$$\frac{d(\beta(R_{0}-1)+\sqrt{\Delta})^{2}}{4A}-d=\frac{d\Delta+d\beta (R_{0}-1)\sqrt{\Delta}}{2A}$$  that
$\frac{d(\beta(R_{0}-1)+\sqrt{\Delta})^{2}}{4A}-d>0$. Then,
$$\begin {array}{lll}
\frac{d(\beta(R_{0}-1)+\sqrt{\Delta})^{2}}{4A}-d>0,\\
\Leftrightarrow d+\frac{k(I_{+}^{*})^{2}}{1+\beta I_{+}^{*}+\alpha (I_{+}^{*})^{2}}-\frac{d(2+\beta I_{+}^{*})}{1+\beta I_{+}^{*}+\alpha (I_{+}^{*})^{2}}-\frac{\mu \delta}{(\delta+d)(\mu+d)}\cdot \frac{k(I_{+}^{*})^{2}}{1+\beta I_{+}^{*}+\alpha (I_{+}^{*})^{2}}>0,\\
\Leftrightarrow a_{3}>0.\\
\end {array}$$
 Clearly, $a_{2}>0,$ and we also have $ a_{1}>0$. By the Routh-Hurartz Criterion, we know that the positive equilibrium $E_{+}^{*}$
is a locally asymptotically stable node.

(ii) For equilibrium $E_{-}^{*},$  we have

$$\begin {array}{lll}
d\alpha (I_{-}^{*})^{2}+k(I_{-}^{*})^{2}-d-\mu \delta \frac{k(I_{-}^{*})^{2}}{(d+\delta)(d+\mu)},\\
=d(\alpha+\frac{k}{d}(1-\frac{\mu\delta}{(d+\delta)(d+\mu)}))(I_{-}^{*})^{2}-d,\\
=d(\alpha+\frac{k}{d}(1-\frac{\mu\delta}{(d+\delta)(d+\mu)}))\frac{(\beta(R_{0}-1)-\sqrt{\Delta})^{2}}{4A^{2}}-d,\\
=\frac{d(\beta(R_{0}-1)-\sqrt{\Delta})^{2}}{4A}-d,\\
<d\frac{2\beta^{2}(R_{0}-1)^{2}-8A-2(\beta^{2}(R_{0}-1)^{2}-4A)}{4A}=0.\\
\end {array}$$
Thus, $a_{3}<0$. By the Routh-Hurartz Criterion, we know in this
case the positive equilibrium $E_{-}^{*}$ is an unstable saddle.
\qed

\begin{table*}[ht]
\caption{The stabilities of the equilibria and the behaviors of
system (1.1) .  }
\begin{center}
\begin{tabular}{|l|lll|l|}
\hline    &$E_{0}$ & $E_{+}^{*}$ &  $E_{-}^{*}$  &System (1.1)
\\\hline
$R_{c}<1$    & LAS &---  &  ---     & Converges to $E_{1}$ \\
 $R_{c}>1$      &LAS &LAS &US   & Bistable\\
\hline
\end{tabular}
\end{center}
\end{table*}

\section{Saddle-node bifurcation}

In this section, we  discuss the bifurcation behavior of system
(1.1). The conditions for saddle-node bifurcation are derived. If
$R_{c}=1$, system (1.1) undergoes a saddle-node bifurcation. The
positive equilibrium $E_{+}^{*}$ and $E_{-}^{*}$ collide to each
other and system (1.1) has a unique instantaneous positive
equilibrium $\bar{E}$. Also one of the eigenvalues of the Jacobian
evaluated at the instantaneous positive equilibrium
  $\bar{E}=(\bar{S},\bar{I},\bar{R})$ is zero. Here
$$\begin {array}{lll}
\bar{S}=\frac{b-(\mu+d-\frac{\delta\mu}{\delta+d})\bar{I}}{d},\\
\bar{I}=\frac{\beta(\mu+d)(R_{0}-1)}{2(\alpha(\mu+d)+\frac{k}{d}(\mu+d-\frac{\delta\mu}{\delta+d}))},\\
\bar{R}=\frac{\mu \bar{I}}{\delta+d}.\\
\end {array}$$

 \noindent{\bf Theorem 4.1} If $R_{c}=1$ or $R_{0}=1+\frac{2}{\beta}\sqrt{\alpha+\frac{k}{d}(1-\frac{
\mu\delta}{(\mu+d)(\delta+d)})}\triangleq R_{0}^{[sn]}$, system
(1.1) undergoes a saddle-node bifurcation around instantaneous
positive equilibrium $\bar{E}=(\bar{S},\bar{I},\bar{R})$.

\noindent{\bf Proof.}

Let  $R_{0}$ be the bifurcation parameter. We use the Sotomayor's
theorem to prove that system (1.1) undergoes a saddle-node
bifurcation. The Jacobian matrix at the saddle-node must have a zero
eigenvalue and two eigenvalues with negative real parts. Let
$F=(f_{1},f_{2},f_{3})^{T}$ with

$$\begin {array}{lll}
f_{1}=b-dS-\frac{kSI^{2}}{1+\beta I+\alpha I^{2}}+\delta R,\\
~~~=b-dS-R_{0}\frac{\beta d(\mu+d)SI^{2}}{b(1+\beta I+\alpha I^{2})}+\delta R,\\
f_{2}=\frac{kSI^{2}}{1+\beta I+\alpha I^{2}}-(\mu+d)I,\\
~~~=R_{0}\frac{\beta d(\mu+d)SI^{2}}{b(1+\beta I+\alpha I^{2})}-(\mu+d)I,\\
f_{3}=\mu I-(\delta+d)R.\\
\end {array}$$

The Jacobian matrix of system  (1.1) at $\bar{E}$ is given by

$$\mathscr{J}_{\bar{E}}=\left[
\begin{array}{cccc}
-d-\frac{k\bar{I}^{2}}{1+\beta \bar{I}+\alpha \bar{I}^{2}}     &-\frac{(\mu+d)(2+\beta \bar{I})}{1+\beta \bar{I}+\alpha \bar{I}^{2}}       &\delta \\
\frac{k\bar{I}^{2}}{1+\beta \bar{I}+\alpha \bar{I}^{2}}        &\frac{(\mu+d)(2+\beta \bar{I})}{1+\beta \bar{I}+\alpha \bar{I}^{2}}-(\mu+d)   & 0  \\
0    &\mu    &-(\delta+d)  \\
\end{array}
\right].$$
%%20191123

The matrix has a simple zero eigenvalue, which requires that
$det(\mathscr{J}_{\bar{E}})=0$ at $R_{0}=R_{0}^{[sn]}$. If $V$ and
$W$ represent eigenvectors corresponding to the eigenvectors of
$\mathscr{J}_{\bar{E}}$ and $\mathscr{J}_{\bar{E}}^{T}$
corresponding to the zero eigenvalue, respectively, then they are
given by

$$V=\left[
\begin{array}{cccc}
v_{1}\\
v_{2}\\
v_{3}  \\
\end{array}
\right]=\left[
\begin{array}{cccc}
\frac{\delta\mu-(\mu+d)(\delta+d)}{\mu d}\\
\frac{\delta+d}{\mu}\\
1  \\
\end{array}
\right],$$

$$W=\left[
\begin{array}{cccc}
w_{1}\\
w_{2}\\
w_{3}  \\
\end{array}
\right]=\left[
\begin{array}{cccc}
1\\
1+\frac{d(1+\beta \bar{I}+\alpha \bar{I}^{2})}{k\bar{I}^{2}}\\
\frac{d(\mu+d)(1-\alpha \bar{I}^{2})+k\bar{I}^{2}(\delta-\mu-d)}{k\bar{I}^{2}(\mu-\delta-d)}  \\
\end{array}
\right].$$ Thus we get

$$F_{c}(\bar{E},R_{0}^{[sn]})=\left[
\begin{array}{cccc}
-\frac{\beta d(\mu+d)\bar{S}\bar{I}^{2}}{b(1+\beta \bar{I}+\alpha \bar{I}^{2})}\\
\frac{\beta d(\mu+d)\bar{S}\bar{I}^{2}}{b(1+\beta \bar{I}+\alpha \bar{I}^{2})}\\
0\\
\end{array}
\right],$$

$$D^{2}F(\bar{E},R_{0}^{[sn]})(v,v,v)=\left[
\begin{array}{cccc}
\frac{(\mu+d)(\beta\alpha \bar{I}^{2}+4\alpha\bar{I}+\beta)}{(1+\beta \bar{I}+\alpha \bar{I}^{2})^{2}}(\frac{\delta+d}{\mu})^{2}-\frac{k\bar{I}(2+\beta \bar{I})}{(1+\beta \bar{I}+\alpha \bar{I}^{2})^{2}}\cdot \frac{\delta\mu-(\mu+d)(d+\delta)}{\mu+d}\cdot\frac{\delta+d}{\mu}\\
-\frac{(\mu+d)(\beta\alpha \bar{I}^{2}+4\alpha\bar{I}+\beta)}{(1+\beta \bar{I}+\alpha \bar{I}^{2})^{2}}(\frac{d+\delta}{\mu})^{2}+\frac{k\bar{I}(2+\beta \bar{I})}{(1+\beta \bar{I}+\alpha \bar{I}^{2})^{2}}\cdot \frac{\delta\mu-(\mu+d)(d+\delta)}{\mu+d}\cdot\frac{\delta+d}{\mu}\\
0\\
\end{array}
\right].$$ Clearly,

$$W^{T}F_{\beta}(\bar{E},R_{0}^{[sn]})=\frac{\beta (\mu+d)\bar{S}d^{2}}{bk} \neq 0,$$

$$\begin {array}{lll}
W^{T}D^{2}F(\bar{E},R_{0}^{[sn]})(v,v,v)=-(\frac{(\mu+d)(\beta\alpha \bar{I}^{2}+4\alpha\bar{I}+\beta)}{(1+\beta \bar{I}+\alpha \bar{I}^{2})^{2}}(\frac{\delta+d}{\mu})^{2}-\frac{k\bar{I}(2+\beta \bar{I})}{(1+\beta \bar{I}+\alpha \bar{I}^{2})^{2}}\cdot \frac{\delta\mu-(\mu+d)(d+\delta)}{\mu+d}\cdot\frac{\delta+d}{\mu})\frac{d(1+\beta \bar{I}+\alpha \bar{I}^{2})}{k\bar{I}^{2}}\neq 0.\\
\end {array}$$
Therefore, from the Sotomayor's theorem, system (1.1) undergoes a
saddle-node bifurcation around instantaneous positive equilibrium
$\bar{E}=(\bar{S},\bar{I},\bar{R})$ at $R_{0}=R_{0}^{[sn]}$. Hence,
we can conclude that when the parameter a passes from one side of
$R_{0}=R_{0}^{[sn]}$ to the other side, the number of positive
equilibria of system (1.1) changes from zero to two.

\section{Numerical simulations and Discussion }

\textbf{5.1 Numerical simulations}

To verify our analytical results,  we carry out some numerical
simulations. In the following, we fix the parameter values as
follows:
$$\hspace{0.02cm}  b=1.5,   d=0.02, k=0.002,
 \delta=0.5, \mu=0.1, \alpha=0.04.
   \eqno(5.1)$$
If we choose $\beta=0.7$, the thresholds  $R_{0}^{[sn]}\approx 1.67$
and $R_{c}\approx 1.09$. In this case, we have a saddle-node
bifurcation (Figure 1).  When $ R_{0}= 2.5$,  two  equilibria of the
model $E_{+}^{*}$ and $E_{0}$ are stable (Figure 2). If we choose
parameter values, such that $R_{0}= 1.25$, then we have only one
equilibrium $E_{0}$ which is stable (Figure 3);

\begin{figure}[!h]
\begin{center}
{\rotatebox{0}{\includegraphics[width=0.7 \textwidth,
height=50mm]{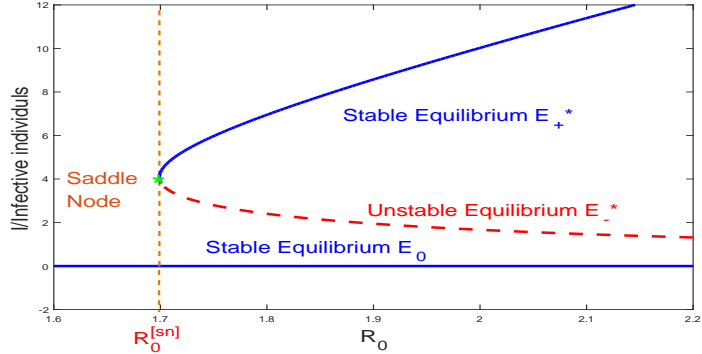}}}
 \caption{
\footnotesize  Bistability and saddle-node bifurcation diagram of
system (1.1). In this case, $R_{0}^{[sn]}\approx 1.67$. The system
displays two stable equilibria $E_{0}$ (the blue solid line at the
bottom) and $E_{+}^{*}$ (the above blue curve), indicating bistable
behaviour. Here, $\bar{E}$ is the saddle point, where the two
equilibria converge and display saddle-node bifurcation. The point
$E_{-}^{*}$ (dashed lines) on the bottom half of the curve is
unstable, and the point $E_{+}^{*}$ (solid line) on the top half of
the curve is stable. Here, $\beta=0.7$ and other parameter values
are listed in $(5.1)$.}\label{F51}
\end{center}
 \end{figure}

\begin{figure}[!h]
\begin{center}
{\rotatebox{0}{\includegraphics[width=0.48 \textwidth,
height=40mm]{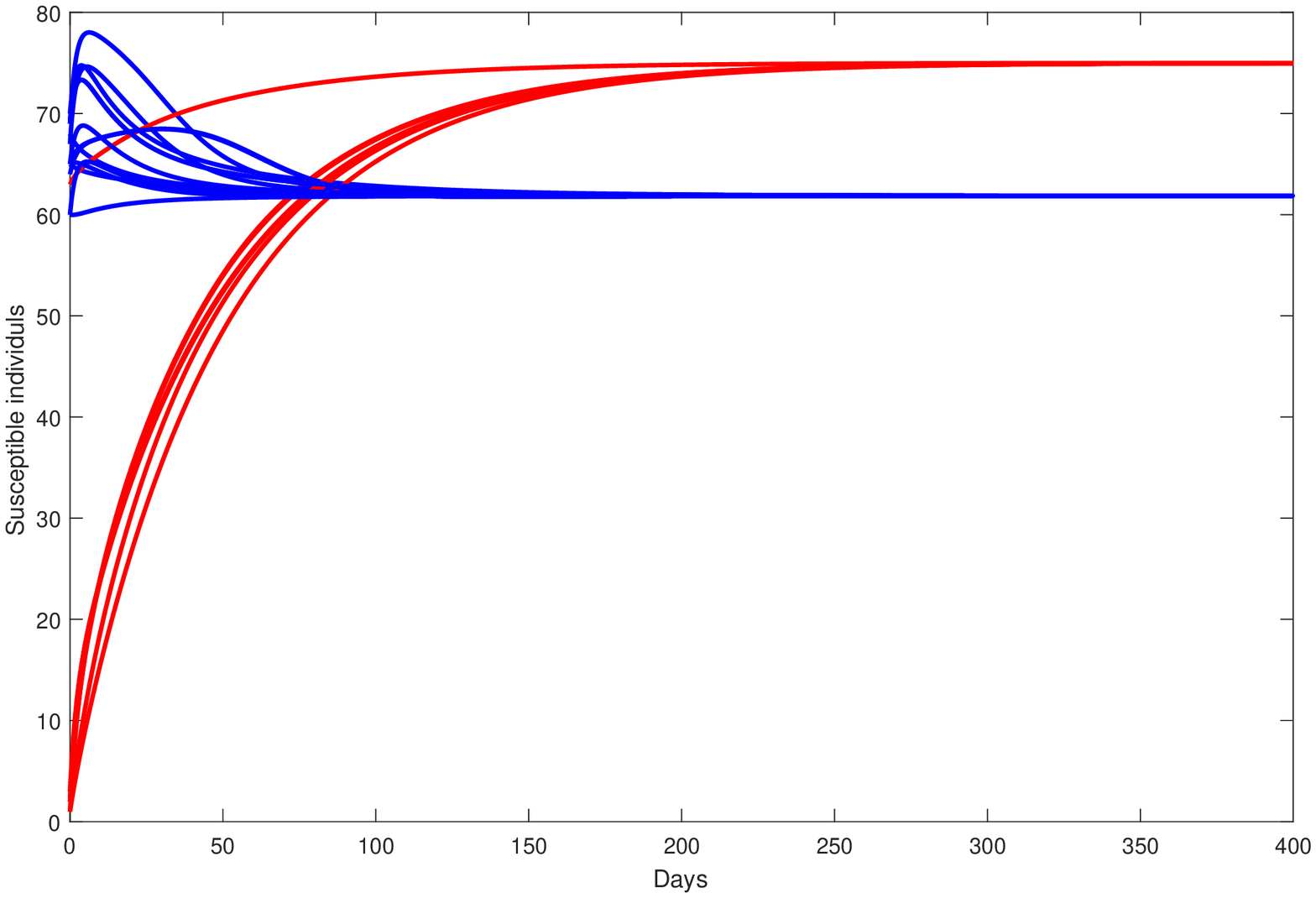}}} {\rotatebox{0}{\includegraphics[width=0.48
\textwidth, height=40mm]{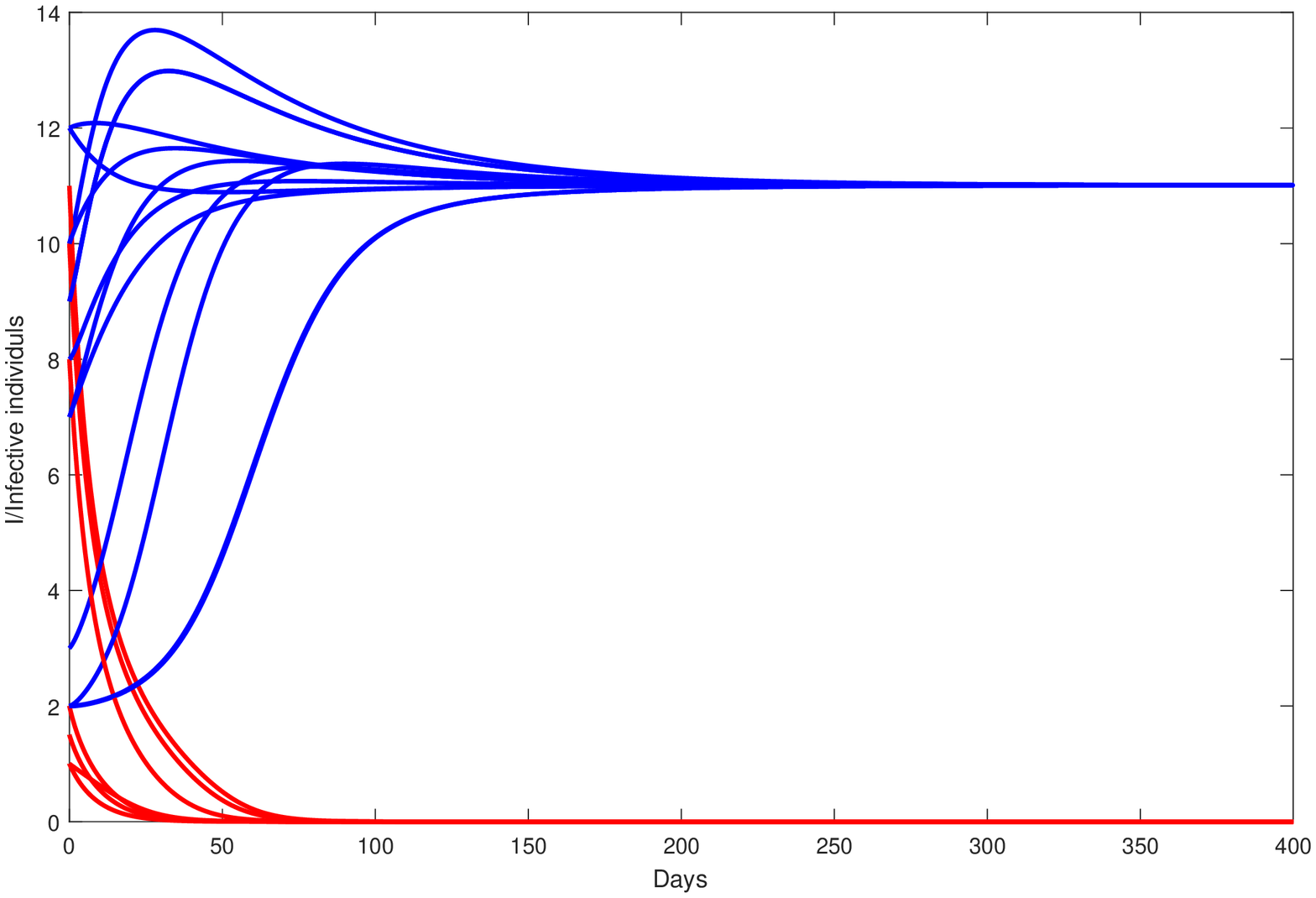}}}
{\rotatebox{0}{\includegraphics[width=0.48 \textwidth,
height=40mm]{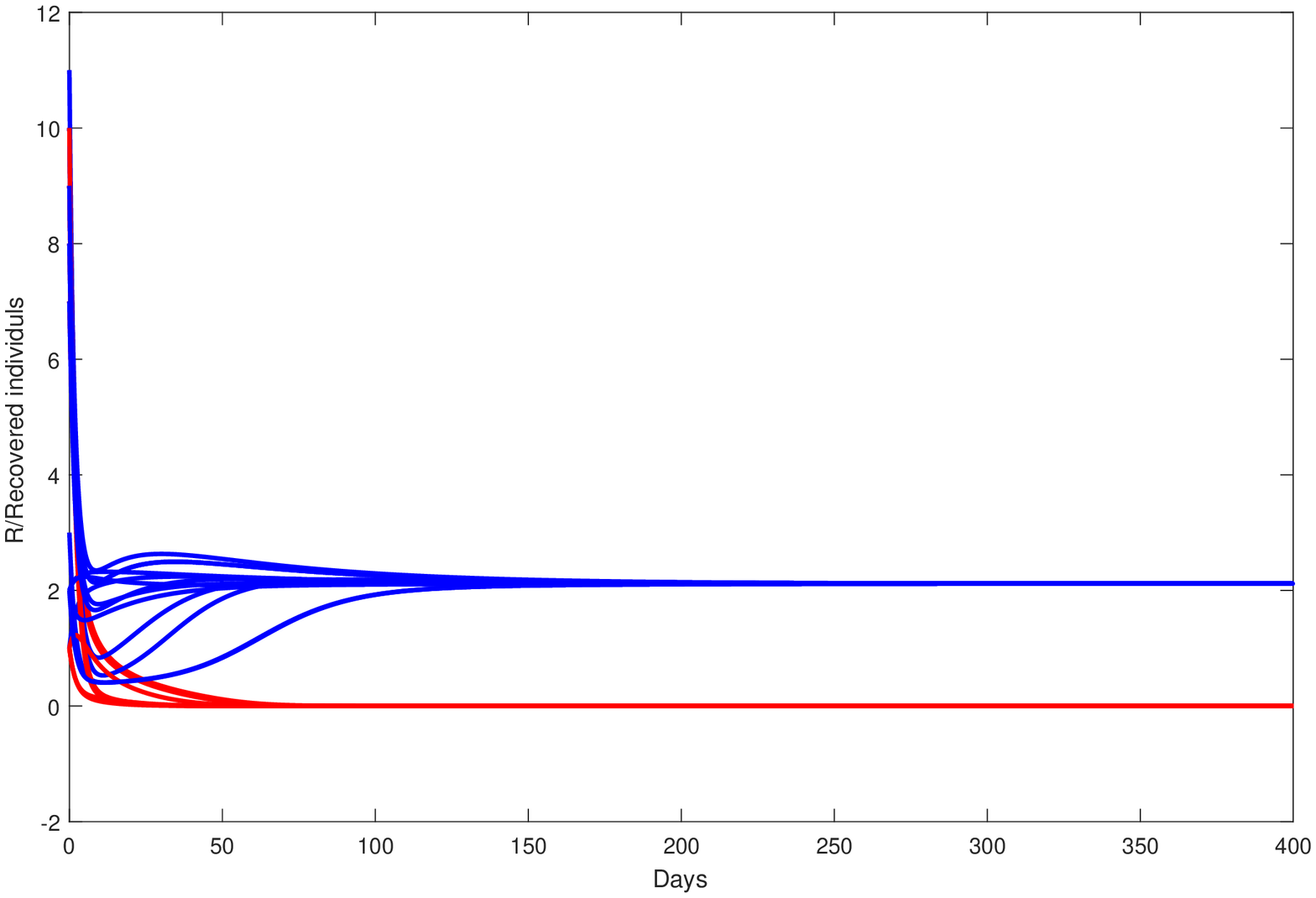}}} {\rotatebox{0}{\includegraphics[width=0.48
\textwidth, height=40mm]{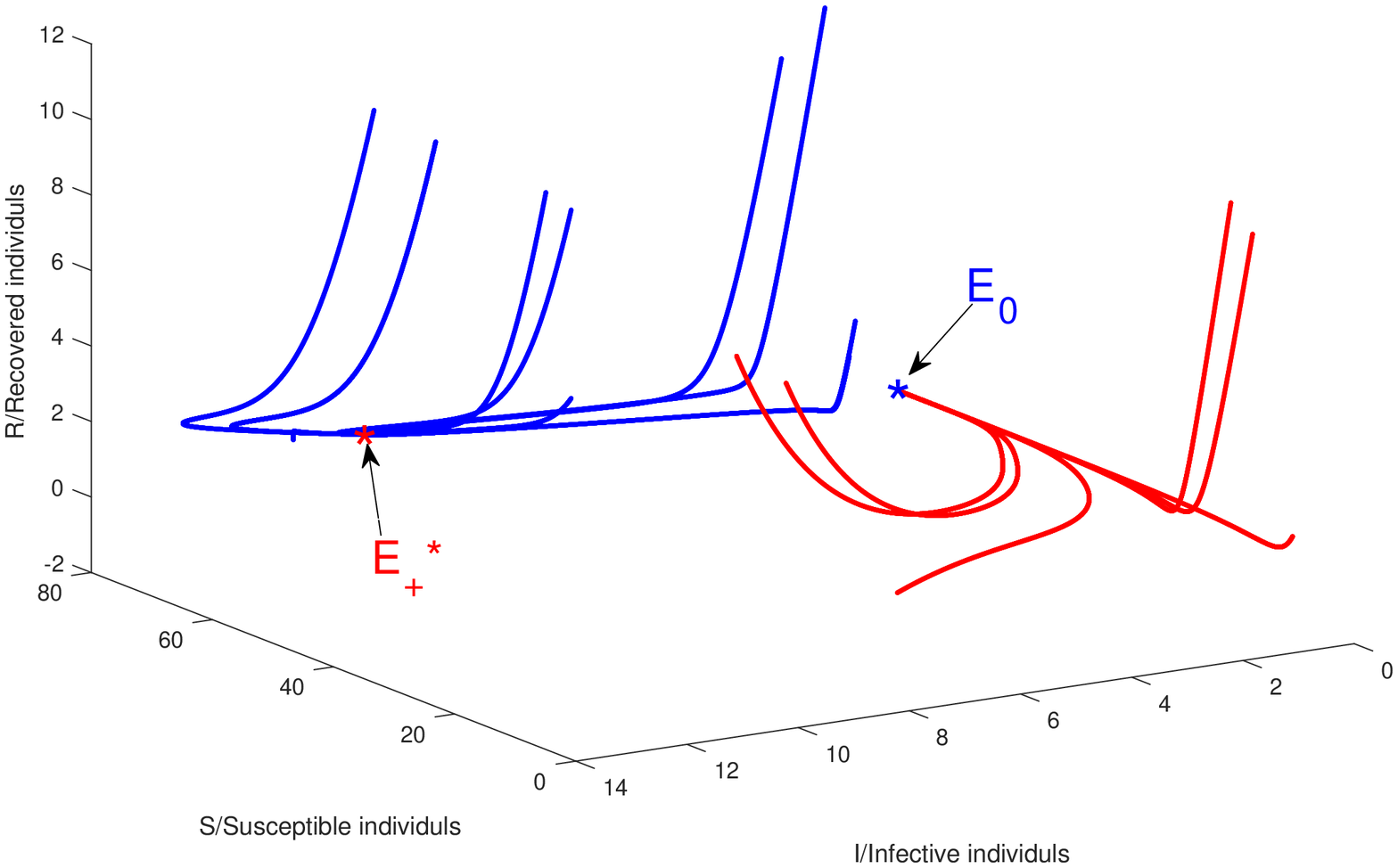}}}
 \caption{
\footnotesize  For $R_{0}=2.5$ and other parameter values listed in
(5.1), we can see that in the case of different initial values, $S$,
$I$ and $R$  converge to either $E_{0}$ or $E_{+}^{*}$.  At this
interval, the system display two stable equilibria $E_{0}$ and
$E_{+}^{*}$, indicating bistable behavior.}\label{F51}
\end{center}
 \end{figure}

%2019 11 29
 \begin{figure}[!h]
\begin{center}
{\rotatebox{0}{\includegraphics[width=0.48 \textwidth,
height=50mm]{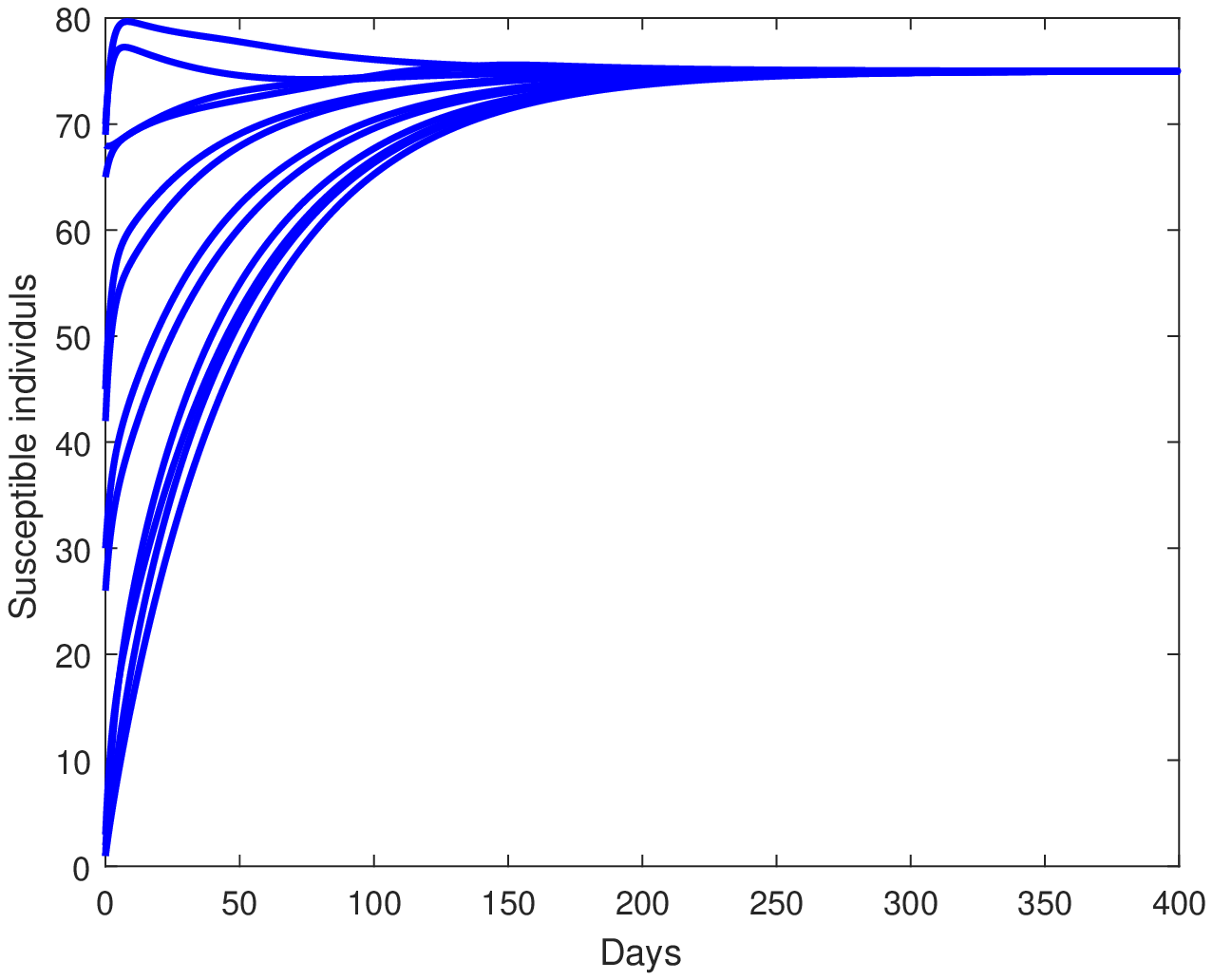}}} {\rotatebox{0}{\includegraphics[width=0.48
\textwidth, height=50mm]{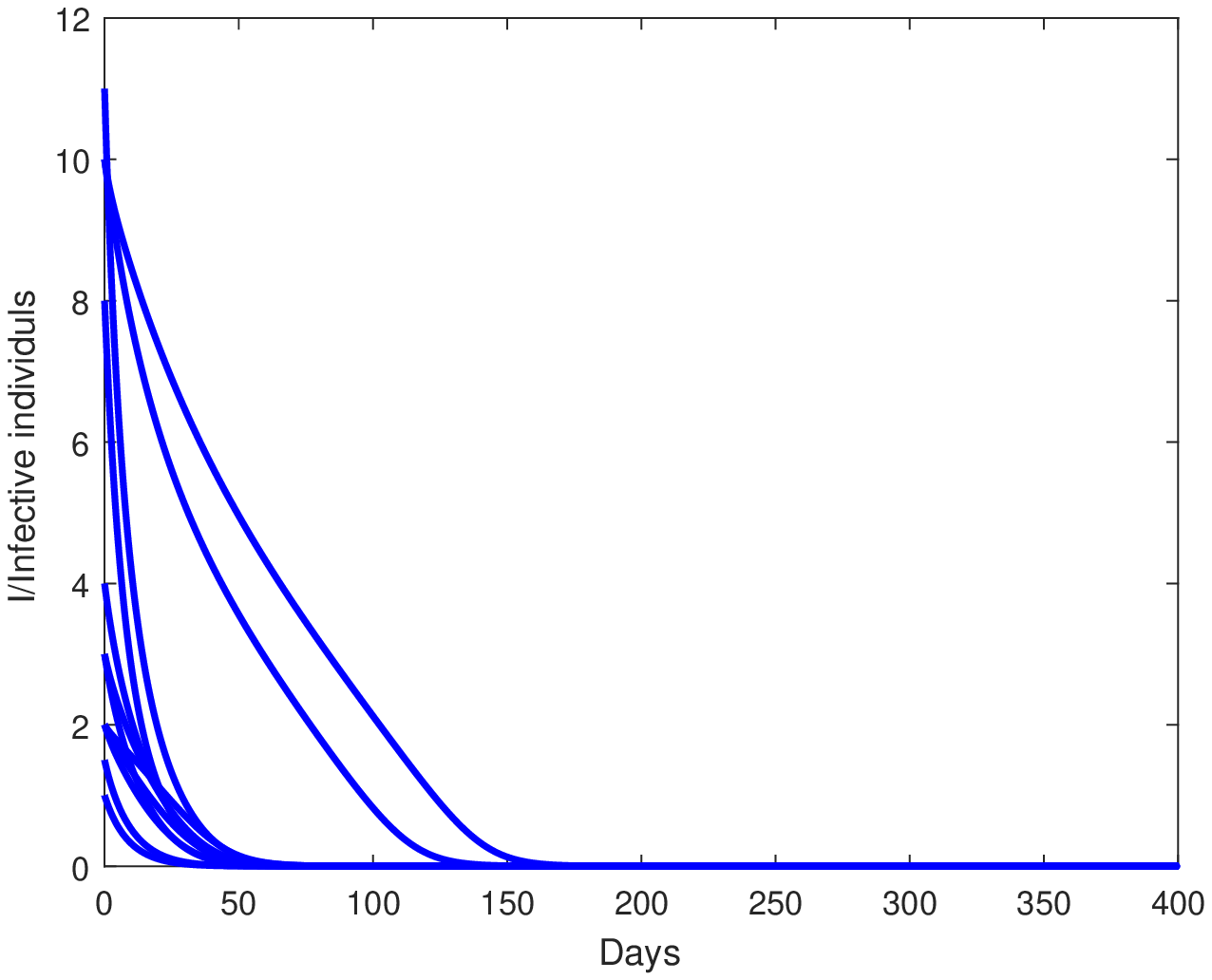}}}
{\rotatebox{0}{\includegraphics[width=0.48 \textwidth,
height=40mm]{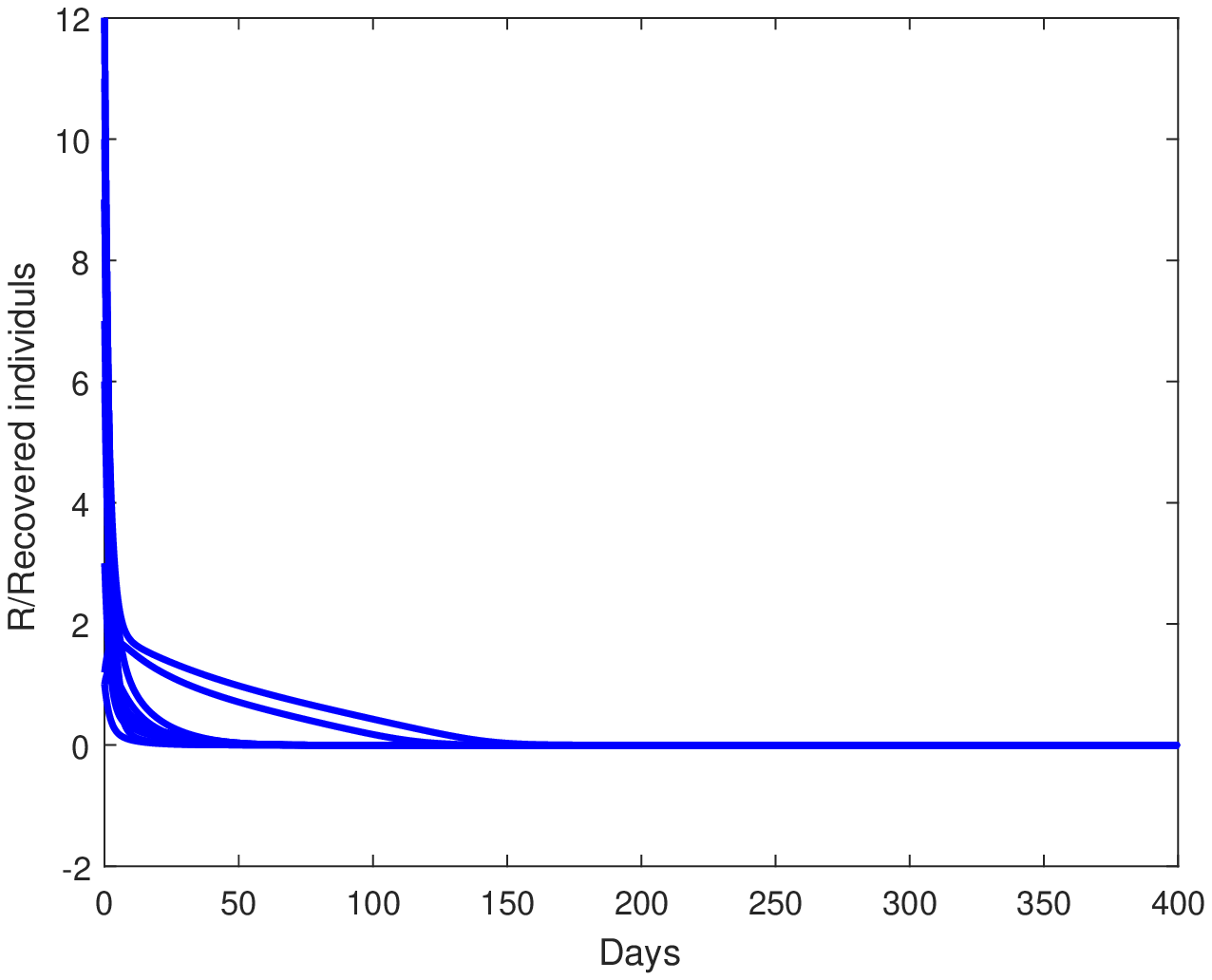}}} {\rotatebox{0}{\includegraphics[width=0.48
\textwidth, height=40mm]{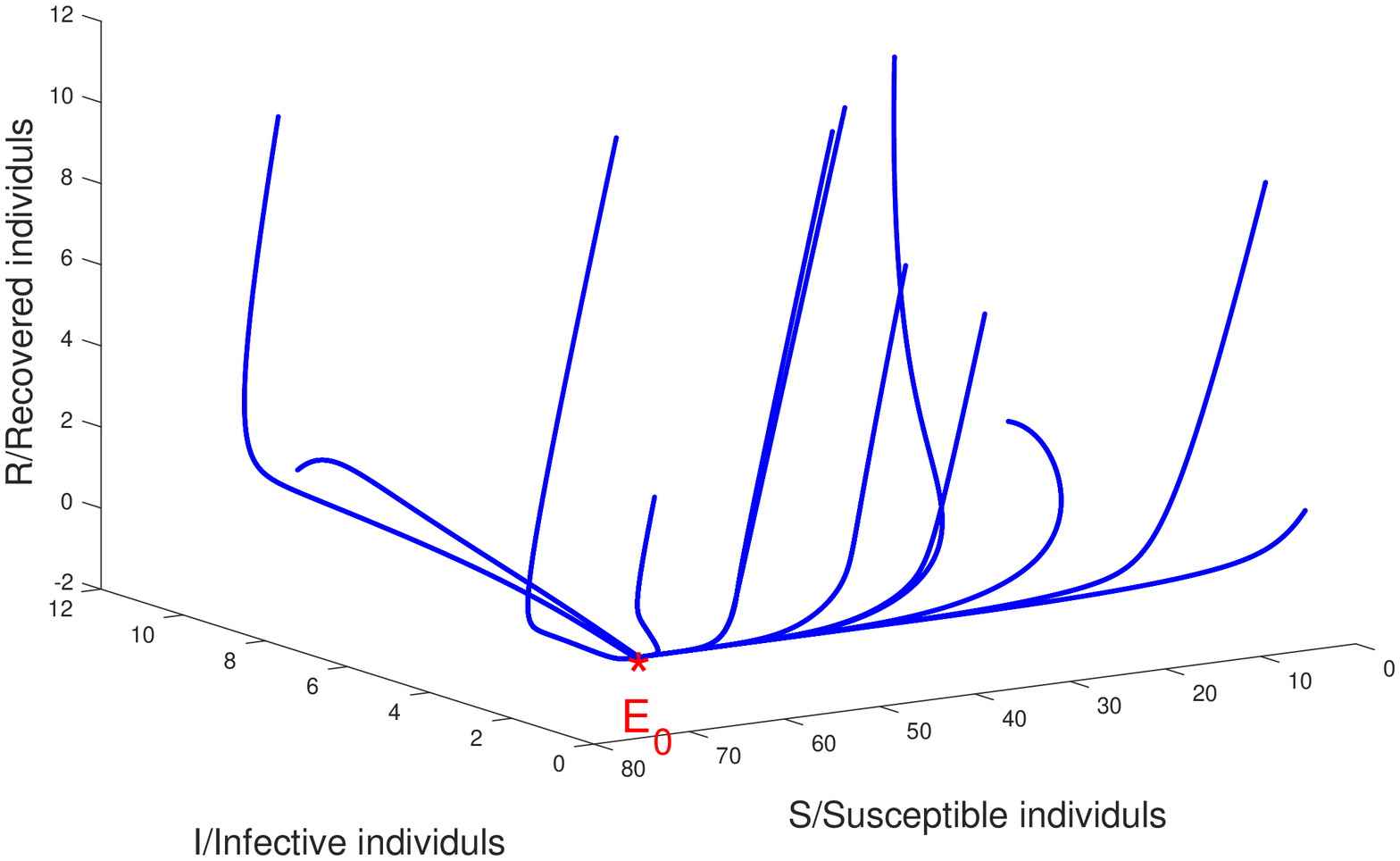}}} \caption{ \footnotesize For
$R_{0}=1.25$ and other parameter values listed in (5.1), we can see
that $S$, $I$ and $R$ converge to $E_{0}$. Here $E_0$  is a locally
asymptotically stable node.}\label{F51}
\end{center}
 \end{figure}

\textbf{5.2 Discussion}

In this paper, we consider a SIRS model with general nonmonotone and
saturated incidence rate and performed mathematical studies. We
found that the system displays bistable behaviours. System (1.1)
admits an disease-free equilibrium $E_{0}$, and two positive
equilibria $E_{+}^{*}$ and $E_{-}^{*}$. We obtain two thresholds
$R_{0}=\frac{bk}{\beta d(\mu+d)}$ and $R_{c}=\frac{bk}{\beta
d(\mu+d)}-\frac{2}{\beta}\sqrt{\alpha+\frac{k}{d}(1-\frac{\mu\delta}{(\mu+d)(\delta+d)})}$.
We find that the system always admits a disease free equilibrium
$E_{0}$ which  is always asymptotically stable, indicating that
there is no infective in the system and all individuals are
susceptible. Thus, if there is no disease, then the uninfected state
will remain stable for a long time. When $R_{c}>1$ , both
$E_{+}^{*}$ and $E_{-}^{*}$ exist, where $E_{+}^{*}$ is locally
asymptotically stable and $E_{-}^{*}$ is unstable, which implies the
coexistence of susceptible, infective and recovered individuals.
Choosing $R_{0}$ as the branching parameter, our investigation
implies that if $R_{c}=1$ or $R_{0}=R_{0}^{[sn]}$ system (1.1)
undergoes a saddle-node bifurcation. The positive equilibria
$E_{+}^{*}$ and $E_{-}^{*}$ collide to each other and system (1.1)
has the unique instantaneous endemic equilibrium $\bar{E}$. From the
branch diagram in figure 1, we find that when $R_{0}>R_{0}^{[sn]}$,
the system has two stable equilibria $E_{+}^{*}$ and $E_{0}$ appear.
The system displays bistable behaviour. When $R_{0}<R_{0}^{[sn]}$,
the system has only one equilibrium point $E_{0}$, suggesting that
infectious diseases will die out eventually.

Castillo-Chavez and Song \cite{Carlos-Castillo-Chavez-Baojun-Song}
proposed the backward bifurcation to illustrate that even if the
basic reproduction number $R_0<1$, disease outbreaks are still
possible. The backward bifurcation indicates that the system
displays bistable behavior when the bifurcation point
$R_{c}<R_0<1$.However, when $R_0 > 1$, the system has only one
positive equilibrium point, which is stable, and the disease-free
equilibrium point is unstable.

In this paper, we investigated a SIRS model with general nonmonotone
and saturated incidence rate. We find that (i) the disease-free
equilibrium is always stable. (ii) When $1<R_0<R_0^{[sn]}$, the
model does not have positive equilibrium point. (iii) When
$R_0>R_{0}^{[sn]}$, the system always display bistability behavior.
Our investigation implies that non-monotonic incidence is a kind of
self-protection behavior of human during the outbreak of a disease.
Such self-protection behavior can reduce the basic reproduction
number and lead to bistable behavior, i.e., there may or may not be
a disease outbreak. Our investigation implies that $R_0$ is not the
basic reproduction number of the model. We guess that $R_{c}$ may be
the basic reproduction number. However, $R_{c}>1$ does not guarantee
the outbreak of the disease, which depends on human behavior. How to
calculate the basic regeneration number of this model is still an
open question.

In other epidemic models with general nonmonotone and saturated
incidence rate \cite{Wang1,Wang2}, we also find such bistable
phenomenon.

\end{document}